\documentclass[usenatbib]{mnras}
\usepackage[english]{babel}	
\usepackage[utf8]{inputenc}
\usepackage{amsfonts}
\usepackage{amsmath}
\usepackage{siunitx}
\usepackage{supertabular}
\usepackage{enumitem}
\usepackage{soul}
\usepackage{xcolor}					
	\definecolor{nblue}{HTML}{8c8cd9}
	\setulcolor{nblue}
	\setul{}{0.1px}
	
\usepackage[T1]{fontenc}
\usepackage{ae,aecompl}

\usepackage{graphicx}	
\usepackage[config]{caption,subfig}
\usepackage{amsmath}	
\usepackage{amssymb}	
\usepackage{tabu}	

\newcommand{\Msun}{{\rm M_\odot}}	
\newcommand{\rhoCrit}{\rho_{\rm crit}}	
\newcommand{\centplus}{\raisebox{.1em}{+}} 

\title[Substructure of Abell 2744]{Abell 2744: Too {\it much} substructure for $\Lambda$CDM?}

\author[J. Schwinn et al.]{%
J. Schwinn$^{1,2}$\thanks{E-mail: johannes.schwinn@stud.uni-heidelberg.de},
M. Jauzac$^{3,1,4}$,
C. M. Baugh$^{1}$,
M. Bartelmann$^{2}$, 
D. Eckert$^{5}$,\newauthor
D. Harvey$^{6}$,
P. Natarajan$^{7}$,
R. Massey$^{3,1}$\\
$^{1}$Institute for Computational Cosmology, Departement of Physics, University of Durham, South Road, Durham DH1 3LE, U.K.\\
$^{2}$Universit\"at Heidelberg, Zentrum f\"ur Astronomie, Institut f\"ur Theoretische Astrophysik, Philosophenweg 12, 69120 Heidelberg, Germany\\
$^{3}$Centre for Extragalactic Astronomy, Department of Physics, Durham University, Durham DH1 3LE, U.K.\\
$^{4}$Astrophysics and Cosmology Research Unit, School of Mathematical Sciences, University of KwaZulu-Natal, Durban 4041, South Africa\\
$^{5}$Astronomy Department, University of Geneva, 16 ch. d’Ecogia, CH-1290 Versoix, Switzerland\\
$^{6}$Laboratoire d’Astrophysique, Ecole Polytechnique F\'ed\'erale de Lausanne (EPFL), Observatoire de Sauverny, CH-1290 Versoix, Switzerland\\
$^{7}$Department of Astronomy, 52 Hillhouse Avenue, Steinbach Hall, Yale University, New Haven, CT 06511, USA
}

\date{Accepted XXX. Received YYY; in original form ZZZ}

\pubyear{2016}

\begin{document}
\label{firstpage}
\pagerange{\pageref{firstpage}--\pageref{lastpage}}
\maketitle

\begin{abstract}
The massive substructures found in Abell 2744 by \cite{Jauzac2016} present a challenge to the 
cold dark matter paradigm due to their number and proximity to the  cluster centre. We use one of 
the biggest N-body simulations, the Millennium XXL, to investigate the substructure in a large sample of 
massive dark matter haloes.  A range of effects which influence the comparison with the observations is 
considered, extending the preliminary evaluation carried out by \cite{Jauzac2016}.  There are many tens of haloes 
in the simulation with a total mass comparable with or larger than that of Abell 2744. However, we find no 
haloes with a number and distribution of massive substructures ($> 5 \times 10^{13} \,{\rm M}_{\odot}$) that is close 
to that inferred from the observations of Abell 2744. The application of extreme value statistics suggests that we would 
need a simulation of at least ten times the volume of the Millennium XXL to find a single dark matter halo 
with a similar internal structure to Abell 2744.  Explaining the distribution of massive substructures in clusters is a 
new hurdle for hierarchical models to negotiate, which is not weakened by appeals to baryonic physics or uncertainty 
over the nature of the dark matter particle.  
\end{abstract}



\begin{table*}
  \caption{The substructures of Abell 2744. Column 1 gives the id of the substructure, column 2 and 3 give the RA and dec respectively, column 4 gives the mass within a circular aperture of radius 150 kpc, column 5 gives the significance level of the detection in units of the variance ($\sigma$) in the mass map and column 6 gives the distance of the substructure from the {\it Core}'s brightest cluster galaxy (BCG). This table is based on table 2 from \citet{Jauzac2016}. Note the BCG has right ascension $\alpha = \ang{3.586259}$ and declination $\delta = \ang{-30.400174}$.}
  \centering
    \begin{tabular}{cllccc}
  \hline\hline
  \noalign{\smallskip}
    {\it ID} & R.A. & Dec. & $M(r < 150 \, {\rm kpc})$ & $\sigma$ & $D_{C-S}$ \\
    &(deg)&(deg)&($10^{13} \Msun$)&&(kpc)\\[0.2em] \hline 
    {\it Core} & 3.586259 & -30.400174 & $13.55\pm 0.09$ & 150& - \\
    {\it N} & 3.5766583 & -30.357592 & $6.10\pm 0.50$ & 12 & 708.4 \\
    {\it NW} & 3.5530963 & -30.376764 & $7.90\pm 0.60$ & 13 & 603.6\\
    ${\it W}_{\it bis}$ & 3.5462875 & -30.403319& $5.20\pm 0.60$ & 9 & 565.3\\
    {\it S1} & 3.6041246 & -30.37465 & $5.00\pm 0.40$ & 13 & 486.9 \\
    {\it S2} & 3.59895 & -30.356925 & $5.40\pm 0.50$ & 11 & 728.5 \\
    {\it S3} & 3.5415083 & -30.373778 & $6.50\pm 0.60$ & 11 & 763.7 \\
    {\it S4} &3.524725 & -30.369583 & $5.50\pm 1.20$ & 5 & 1000.5 \\
    \hline\hline
  \end{tabular}
\label{tab:substructures}
\end{table*}%

\section{Introduction}

The cold dark matter plus cosmological constant model ($\Lambda$CDM) is now established as the standard model of cosmology. This model describes many observations remarkably well, such as the fluctuations in the cosmic microwave background \citep{Planck2015}, the accelerated expansion of the Universe as inferred from the Hubble diagram of type-Ia supernovae \citep{Riess1998, Perlmutter1999} and the large-scale clustering of galaxies \citep[e.g.][]{Cole2005,Alam2016}. Much recent attention has been focused on small scale challenges to the model (for an overview see \citealt{Weinberg2015}). The resolution of these issues typically requires the consideration of baryonic physics (e.g. \citealt{Sawala2014}). Here we consider a new test of $\Lambda$CDM which considers the properties of massive substructures in cluster-mass dark matter haloes. This has the attraction that the substructures are so massive that baryonic physics \citep{Munari2016} and the nature of the dark matter particle (for example, allowing interactions between the dark matter and standard model particles or dark matter self-interactions, see \citealt{Boehm2014,Robertson2016}) have been argued to be unimportant.  

Observations of large galaxy clusters at intermediate redshifts have often been put forward as a challenge to models in which structure forms hierarchically, since in this case the most massive haloes form more recently than low mass haloes \citep[e.g.][]{Broadhurst2008,Jee2009,Foley2011,Hoyle2011,Holz2012}. This tension was resolved by \cite{Hotchkiss2011} who pointed out that the way in which the observation of a single cluster at a given redshift is compared with the theoretical model has a huge impact on the probability of finding such objects. By framing the question in a more appropriate way, namely by calculating the chance of finding an object at least as massive as the cluster at the observed redshift or higher, these objects can be shown to be compatible with $\Lambda$CDM. An alternative but related approach was introduced by \cite{Davis2011} and \cite{Waizmann2011}, who used extreme value statistics to predict the probability of finding the most massive galaxy clusters in a $\Lambda$CDM universe. These authors also do not find a conflict with $\Lambda$CDM. 

Here we consider a new way of using massive clusters to constrain hierarchical models \citep[see also][]{Munari2016}. Using both weak and strong gravitational lensing measurements inferred from extensive imaging of Abell 2744 with the Hubble Space Telescope, \citet{Jauzac2016} produced an improved mass reconstruction of the cluster which revealed a remarkably rich degree of substructure. Eight massive substructures (with mass $> 5 \times 10^{13}\,\Msun$) were reported within $\sim 1\,{\rm Mpc}$ of the cluster center. In \cite{Jauzac2016} we compared Abell 2744 with massive dark matter haloes in the Millennium-XXL N-body simulation \citep{Angulo2012}. Although many dark matter haloes of a mass similar to Abell 2744 can be found in the MXXL, none of these contained a similar number of massive subhaloes close to the halo centre (i.e. within a radius of 1\,Mpc). Here we carry out a thorough investigation of various effects which can influence this comparison, extending the preliminary analysis carried out by \cite{Jauzac2016}. In particular we consider: (1) the aperture in which masses are estimated in the lensing analysis, (2) the change in the best-fitting cosmological parameters since the MXXL simulation was run and (3) the Eddington bias which results from errors in the masses inferred from observations.

Previous tests of the $\Lambda$CDM subhalo mass function have focused on somewhat smaller masses than we consider here, $10^{10}$ -- $10^{12.5} \, {\rm M}_{\odot}$, and reported agreement with observations of lensing clusters \citep[][]{Natarajan2004, Natarajan2007}. On the other hand \cite{Grillo2015} found an excess of subhaloes with effective circular velocities in the range $100$--$300 \,{\rm km \,s}^{-1}$ in another \emph{Hubble Frontier Fields} \citep[HFF, ][]{Lotz2016} cluster MACS J0416 compared to clusters of a dark matter only simulation. Also these subhaloes are of lower masses than the ones considered in this paper and thus baryonic physics might influence their abundance. A more recent comparison of the subhalo mass function in the inner regions of the 3 HFF clusters on lower mass scales than the present work (10$^9$--10$^{12.5}\,\Msun$), using resimulations of clusters with galaxy formation treated using the Illustris simulations \citep[][]{Vogelsberger2014} and analytic calculations find excellent agreement between observations and the predictions of $\Lambda$CDM (Natarajan et al. submitted MNRAS). However, these authors do report that the observed radial distribution of substructure does not match the distribution seen in the Illustris clusters. Natarajan et al. claim that this discrepancy arises since exact dynamical analogues of the HFF clusters, even if they are mass matched with simulated ones, are not available in current simulation volumes. In contrast to Natarajan et al., we examine here the radial distribution of the rarer more massive substructures.

The paper is structured as follows. In Section~\ref{sec:background} we briefly describe the observations of Abell 2744 and give an overview of the MXXL simulation. We compare halo masses derived from simulations with those inferred from observations in Section~\ref{sec:massSimuObs}. The adjustments made to the halo masses in MXXL to compare to observations are presented in Section~\ref{sec:massMXXLCorr}. We describe the search for haloes similar to Abell 2744 in Section~\ref{sec:searchMXXL}. The tidal stripping of MXXL subhaloes is discussed in Section~\ref{sec:tidalStripping}. In Section~\ref{sec:extremeVal}, we use extreme value statistics to estimate the probability of finding haloes like Abell 2744 in a $\Lambda$CDM universe. We conclude in Section~\ref{sec:summary}.

\section{Observational and simulated datasets}
\label{sec:background}

Below we summarize recent observations of Abell 2744 and their interpretation in Section~\ref{subsec:background_A2744} before introducing the Millennium XXL simulation in Section~\ref{subsec:background_MXXL}.

\subsection{Abell 2744}
\label{subsec:background_A2744}
Abell 2744 is one of the most massive and complex galaxy clusters known, with a total mass of $\sim 3 \times 10^{15}\,\Msun$ at redshift $z=0.308$. Its rich structure has been highlighted in a series of papers. \cite{Merten2011} presented the first strong and weak-lensing analyses of Abell 2744 and combined these results with {\it Chandra} X-ray observations from \cite{Owers2011}. This revealed four massive substructures in the core of the cluster, all with masses typical of cluster-mass haloes ($\sim$10$^{14}$\,M$_{\odot}$): the {\it Core}, the {\it Northern} (N), the {\it North-Western} (NW) and the {\it Western} (W) clumps. \cite{Medezinski2016} took advantage of new (non-public) weak-lensing data from Subaru to analyse the mass distribution in the outskirts of the cluster and compared their results to those of \cite{Merten2011}. \cite{Medezinski2016} detected a fifth cluster-mass substructure in the North-East of the {\it Core}.

Given its special nature, Abell 2744 was selected as one of the HFF \citep{Lotz2016} clusters. It has been extensively observed with the {\it Hubble Space Telescope} (HST) with both the {\it Advanced Camera for Surveys} (ACS) and the {\it Wide-Field Camera 3} (WFC3) for a total of 140 orbits in 7 pass-bands from the optical to the near-infrared. These observations led the lensing community to revise the mass model of the {\it Core} component of the cluster thanks to identifications in more than $180$ multiple images \citep{Jauzac2015,Wang2015}. This allows the mass of the {\it Core} to be constrained to a precision of better than $1$\%. Taking advantage of these new strong-lensing constraints, \cite{Jauzac2016} built a mass model that combines strong and weak-lensing shape measurements from HST and the {\it Canada-France-Hawaii Telescope} (CFHT), allowing the mass distribution to be traced out to $\sim 15\, {\rm Mpc}$ from the cluster {\it Core} \citep[][]{Eckert2015}. \cite{Jauzac2016} increased the mass map resolution compared to that used by \cite{Eckert2015} and investigated the mass distribution within $2\,{\rm Mpc}$ of the {\it Core}. \cite{Jauzac2016} detected eight substructures including all of the substructures from \cite{Merten2011} and \cite{Medezinski2016} plus 3 new ones. All of the substructures are at least $5\,\sigma$ features in the mass maps and have masses ranging from $0.5$ to $ 1.4 \times 10^{14} \, {\rm M}_{\odot}$ (full details of the noise estimation in the mass maps can be found in \citealt{Jauzac2016}). The properties of the substructures are listed in Table~\ref{tab:substructures}. The mass of the total cluster was determined within an aperture of 
1.3\,{\rm Mpc} as $M(R<1.3\,{\rm Mpc}) = \left( 2.3 \pm 0.1 \right) \times 10^{15}\,{\rm M}_\odot$.

Taking advantage of the new X-ray datasets obtained with the {\it XMM-Newton Observatory} in December 2014 \citep{Eckert2015}, \cite{Jauzac2016} reported new detections of one remnant core and one putative shock. The latter, if confirmed, would validate the hypothesis  of \cite{Owers2011} that the North-South axis is the main merger axis of the cluster. Nevertheless, it is extremely difficult to give a definitive dynamical scenario of Abell 2744, if not impossible, considering the complexity of the object.

\subsection{The Millennium XXL Simulation}
\label{subsec:background_MXXL}
The MXXL simulation \citep[][]{Angulo2012} models structure formation in a $\Lambda$CDM universe using 303 billion particles of mass $m_{p} = 8.80 \times 10^{9}\, {\rm M}_{\odot}$ in a cube of side length $3\,h^{-1}{\rm Gpc}$, where $h$ is the Hubble constant today defined in terms of $H_0 = 100h\, \rm{km\,s}^{-1}\rm{Mpc}^{-1}$. The cosmological parameters were set to: $H_0~=~73\,{\rm km\,s^{-1}Mpc^{-1}}$, $\Omega_\Lambda = 0.75$, $\Omega_{\rm m} = \Omega_{\rm dm} + \Omega_{\rm b} = 0.25$, $\Omega_{\rm b} = 0.045$ and  $\sigma_8 = 0.9$. These parameters were chosen to the same as those used in the previous Millennium series runs \citep[][]{Springel2005,Boylan-Kolchin2009}. The cosmological parameters used in MXXL do not correspond to those found by a particular cosmic microwave background experiment; however, they yield a power spectrum of density fluctuations that agrees with that obtained using the \cite{Planck2015} parameters to within 10\%. 

Gravitationally bound structures were identified in the simulation on two different levels.
Dark matter haloes were found using the Friends-of-Friends (FoF) algorithm \citep[][]{Davis1985} and within these FoF haloes gravitationally bound substructures were identified using {\tt SUBFIND} \citep[][]{Springel2001}. The FoF algorithm finds objects whose particle separations are less than a given linking length, $b$. The linking length was set to $b=0.2$, which ensures that FoF haloes enclose an average overdensity of $\sim 180$ times the critical density \citep[see][]{More2011}. The {\tt SUBFIND} algorithm identifies self-bound substructures within the FoF haloes by detecting a saddle point in the density profile. 

Since the volume of the MXXL is more than 10 times the volume contained in the whole sky out to redshift $z~=~0.308$, and given the ability to resolve substructure in the most massive haloes, the simulation provides a unique opportunity to analyse the frequency and structure of rare objects like the Abell 2744 cluster in a $\Lambda$CDM universe. We performed our search on the MXXL snapshot at redshift $z = 0.32$ and the subsequent snapshot at $z = 0.28$, which bracket the redshift of Abell 2744 ($z=0.308$).

\section{Comparing halo masses from simulations with observations}
\label{sec:massSimuObs}

\subsection{The mass of dark matter haloes - definitions}
\label{subsec:background_theory}
A typical definition used for the mass of a dark matter halo is the mass enclosed within the virial radius. The virial radius is usually approximated by $R_{200}$, which is the radius enclosing a mean overdensity of 200 times the critical density of the universe $\rhoCrit$. The mass within a sphere of this radius is 
\begin{equation}
M_{200} = 200 \, \rhoCrit \, \frac{4}{3} \, \pi \, R_{200}^3\, .
\end{equation}
The Navarro-Frenk-White (NFW) density profile \citep{Navarro1996} is commonly used to describe the radial density distribution of dark matter haloes
\begin{equation}
\rho_{\rm NFW} = \frac{\delta_c \rhoCrit}{x(1+x)^2}\, ,
\end{equation}
where $\delta_c$ is a characteristic overdensity relative to the critical density and $x = r/r_{\rm s}$ is a dimensionless radius, defined in terms of the scale radius $r_{\rm s}$. Using $r_{200}$ and the scale radius, the concentration $c$ of a halo is defined as $c = r_{200}/r_{\rm s}$.

To make our results comparable to those presented in \cite{Jauzac2016}, we will absorb factors of $h$ into the units. Following \cite{Jauzac2016} we adopt a value of $h = 0.7$.

\subsection{Halo masses in the MXXL simulation}
\label{subsec:massMXXL}
\begin{figure}
	\includegraphics[width=\columnwidth]{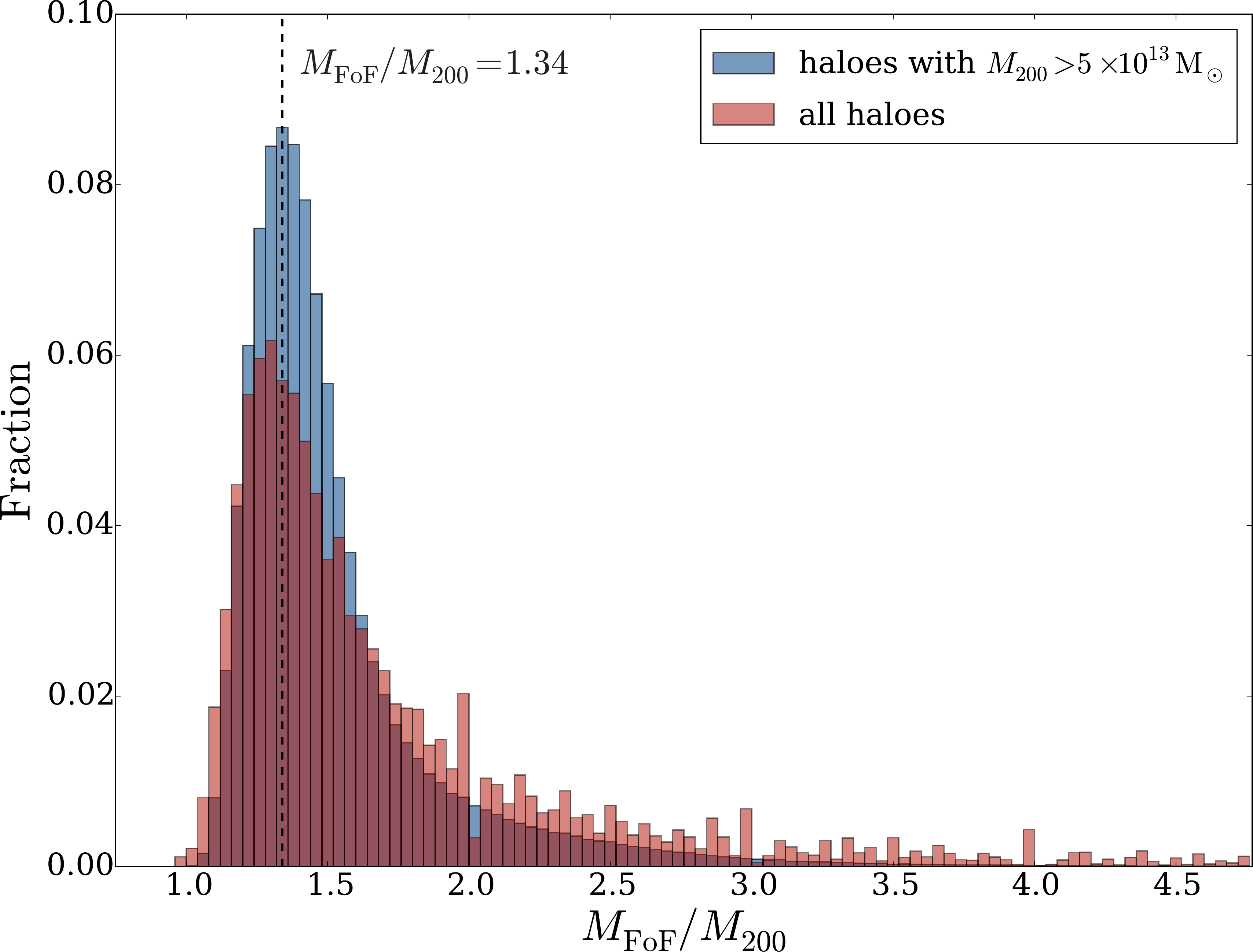}
    \caption{The normalised distribution of the ratio $M_{\rm FoF}/M_{200}$ for haloes with $M_{200} > 5 \times 10^{13}\,\Msun$ (blue) and for the whole population of FoF haloes in MXXL (red), both at $z = 0$.}
    \label{fig:FOFvsM200}
\end{figure}%
The MXXL simulation records four different masses for each FoF halo: $M_{200}$ relative to the critical density at the epoch the halo is detected ($M_{200,\rm crit}$), $M_{200}$ relative to the mean cosmic matter density ($M_{200,\rm mean}$), the mass within a sphere with density from top-hat collapse model ($M_{\rm tophat}$) and the summed mass of all particles assigned to the particular FoF halo ($M_{\rm FoF}$). For subhaloes identified by {\tt SUBFIND}, only the summed mass of all particles assigned to the subhalo ($M_{\rm sub}$) is stored. Throughout we will use only $M_{\rm FoF}$, $M_{\rm sub}$ and $M_{200, \rm crit}$, so we refer to the latter as $M_{200}$. 

In some instances we will need to convert one halo mass definition into another so we now explore the relationship between $M_{\rm FoF}$ and $M_{200}$. These halo mass estimates often display significantly different values. \citet{Jiang2014} found that the FoF~masses are typically 20-25\% higher than the $M_{200}$ masses for haloes in the mass range $M_{\rm FoF} = 10^{9} - 10^{12}\,\Msun$; this figure exceeds 50\% for haloes with $M_{\rm FoF} = 5\times10^{13}\, \Msun$. Such large deviations between $M_{\rm FoF}$ and $M_{200}$ are in part related to the fact that 15-20\% of FoF haloes have irregular morphologies and can be considered as ``bridged'' haloes \citep[]{Lukic2009}.

Fig.~\ref{fig:FOFvsM200} shows the distribution of the ratio $M_{\rm FoF}/M_{200}$ for all haloes in the MXXL simulation and for the subset of haloes with $M_{200} > 5\times10^{13}\,\Msun$. We find the same result as \citet{Jiang2014} when we calculate the median of the distribution for both halo samples. Both distributions show a sharp peak, which occurs at a ratio of 1.30 for the full halo population and at 1.34 for the high mass haloes. Both distributions show a long tail to high ratios. The more massive haloes tend to have higher $M_{\rm FoF}/M_{200}$ ratios, since the probability that a halo is bridged increases with mass. There are small differences in the tails of the distribution for two reasons: (1) the full halo sample contains more objects, so can throw up more examples of unusual structures and (2) the threshold applied to $M_{200}$ for the high mass halo sample truncates the distribution of $M_{\rm FoF}/M_{200}$, precluding large values of this ratio.

Due to the extended tail of the distribution in Fig.~\ref{fig:FOFvsM200}, we decided not to use the median of the distribution was as done by \citet{Jiang2014}, but instead use the mode. We therefore assume that for haloes with mass $M_{200} > 5\times10^{13}\,\Msun$ the FoF-mass can be estimated by increasing $M_{200}$ by 34\%.

\subsection{Extrapolating aperture masses}
The masses extracted from the lensing reconstructions of Abell 2744 correspond to 
the projected mass in a circular aperture whereas the masses of haloes in N-body 
simulations are measured within a sphere radius of $R_{200}$. The 
circular apertures used by \cite{Jauzac2016} are typically smaller than the $R_{200}$ 
of the objects in question, which means that the mass contained with a sphere of 
radius $R_{200}$ is larger than the mass reported in the circular aperture. 
In this subsection we calculate the correction to the observationally inferred 
masses to obtain a mass that can be compared to the mass of haloes in the 
N-body simulation.

\begin{table}
  \caption{Comparison of mass estimates obtained within apertures of $150 \,{\rm kpc}$ and $250 \,{\rm kpc}$ and the deduced $M_{\rm sub}$ mass for all eight substructures.}
\centering
    \begin{tabular}{cccc}
  \hline\hline\noalign{\smallskip}
    {\it ID} & $M ( r < 150 \, {\rm kpc} ) $ & $M ( r < 250 \, {\rm kpc}) $ & $M_{\rm sub}$ \\\noalign{\smallskip}  
             & ($10^{13} {\rm M}_\odot$) & ($10^{13} {\rm M}_\odot$)  & ($10^{13} {\rm M}_\odot$) \\\noalign{\smallskip} \hline 
    \noalign{\smallskip} {\it Core} & $13.55\pm 0.09$ & $27.7 \pm 0.1$ & $334^{+5}_{-5}$ \\[0.5em]
    {\it N} & $6.10\pm 0.50$ & $14.7 \pm 0.9$ & $58^{+11}_{-9}$\\[0.5em]
    {\it NW} & $7.90\pm 0.60$ & $18.0 \pm 1.0$& $100^{+17}_{-16}$\\[0.5em]
    ${\it W}_{\it bis}$ & $5.20\pm 0.60$ & $12.9 \pm 1.1$ & $43^{+9}_{-9}$\\[0.5em]
    {\it S1} & $5.00\pm 0.40$ & $13.0 \pm 1.0$ & $39^{+7}_{-5}$\\[0.5em]
    {\it S2} & $5.40\pm 0.50$ & - & $46^{+9}_{-8}$ \\[0.5em]
    {\it S3} & $6.50\pm 0.60$ & - & $66^{+14}_{-11}$\\[0.5em]
    {\it S4} & $5.50\pm 1.20$ & - & $47^{+24}_{-17}$ \\[0.5em]
    \hline\hline
  \end{tabular}
\label{tab:extrapolated_masses}
\end{table}%
Since the original particle data is no longer available for MXXL snapshots 52 and 53, it is impossible to obtain directly an aperture mass similar to that provided by the observational analysis. We thus extrapolate the aperture masses of Abell 2744 and its substructures to the masses the corresponding haloes have in the MXXL. To take into account the line-of-sight projection of mass, we assume that the observational mass estimate corresponds to a cylinder whose axis coincides with the centre of the dark matter halo. Therefore, we perform the extrapolation by integrating an NFW density profile over a cylinder corresponding to the properties of the observational aperture. We adopt a radius $R = 1.3\,{\rm Mpc}$ and length $l = 30\,{\rm Mpc}$\footnote{$ R = 1.3\,{\rm Mpc}$ corresponds to the aperture used in \cite{Jauzac2016}. A cylinder length of $l = 30\,{\rm Mpc}$ was chosen because $\rho_{\rm NFW}$ drops below the mean matter density at a radius of $\sim 15\,{\rm Mpc}$.}. The $M_{200}$ mass is obtained by requiring that the integration of the NFW profile over a cylindrical volume gives the mass quoted by \cite{Jauzac2016} (i.e. $M(R<1.3\,{\rm Mpc}) = \left( 2.3 \pm 0.1 \right) \times 10^{15}\,{\rm M}_\odot$). This extrapolation provides the $M_{200}$ mass  We find $M_{200} = 3.3 \pm 0.2 \times 10^{15}\, \Msun$ for the total halo which is 40\% larger than the mass observed within an aperture of $R = 1.3\,{\rm Mpc}$. This result is fairly insensitive to the choice of concentration-mass relation. Adopting the $c$-$M_{200}$ relation from \citet{Neto2007} (our default choice) gives a $M_{200}$ value that is $\sim 10\%$ lower than the mass obtained when using the relation from \cite{Maccio2008}. 

The $M_{200}$ mass of subhaloes is obtained by following a similar procedure, in this case performing the integral for an aperture of $R = 150 \,{\rm kpc}$. This rather small aperture was chosen in \cite{Jauzac2016} to minimise overlap of close subhaloes when determining the substructure masses. To obtain the total mass $M_{\rm sub}$ for such a subhalo, we assume that the $M_{200}$ value has to be increased by 34\% (see Section~3.2). The resulting  masses are shown in Table~\ref{tab:extrapolated_masses}, together with the estimates derived in the gravitational lensing analysis of \cite{Jauzac2016}. In all cases the estimated {\tt SUBFIND} mass is around one order of magnitude higher than the mass contained within the aperture of $150 \,{\rm kpc}$. For the {\it Core} halo in particular, the full subhalo mass is more than 25 times larger than the mass estimate from the lensing analysis, because there is a bigger correction to make in this case going from the cylinder radius to $R_{200}$.

Line-of-sight projection of distinct haloes can also lead to an increase in the estimated mass.  We do not include this effect in our analysis directly, but estimate its impact. We select all haloes with a mass similar to Abell 2744 and place a cylinder with the same size as that used for extrapolating the aperture masses on the centre of each subhalo. We choose for each cylinder a thousand different random orientations relative to the cluster-mass halo and add up the masses of all subhaloes that lie within the cylinder. We find that maximally one subhalo per halo is scattered to a mass higher than $10^{14}\,\Msun$ after accounting for projection effects in this way. However, this event is extremely rare (on average 0.01 subhalo per halo) and thus we do not include this effect in the further analysis.

\section{Adjustments to halo and subhalo masses in the MXXL simulation}
\label{sec:massMXXLCorr}
We now investigate two further effects, beyond the issue discussed in the previous section, that need to be taken into account to make the comparison of masses obtained from observations with those in the MXXL simulation: (i) the change in the current best-fitting cosmological parameters relative to those used in the MXXL and (ii) the Eddington bias due to errors in the observational mass measurements.

\subsection{Correction to account for change in cosmological parameters}
As mentioned above, the MXXL simulation uses cosmological parameters that have been superseded by the Planck measurements \citep{Planck2015}. These changes affect the halo population in two ways: (1) the halo mass function depends on cosmology, so the abundance of objects with a given mass will change and (2) the merger rate of haloes changes. These changes should be taken into account if the MXXL haloes are to be compared to observations.

\begin{figure}
	\includegraphics[width=0.99\columnwidth]{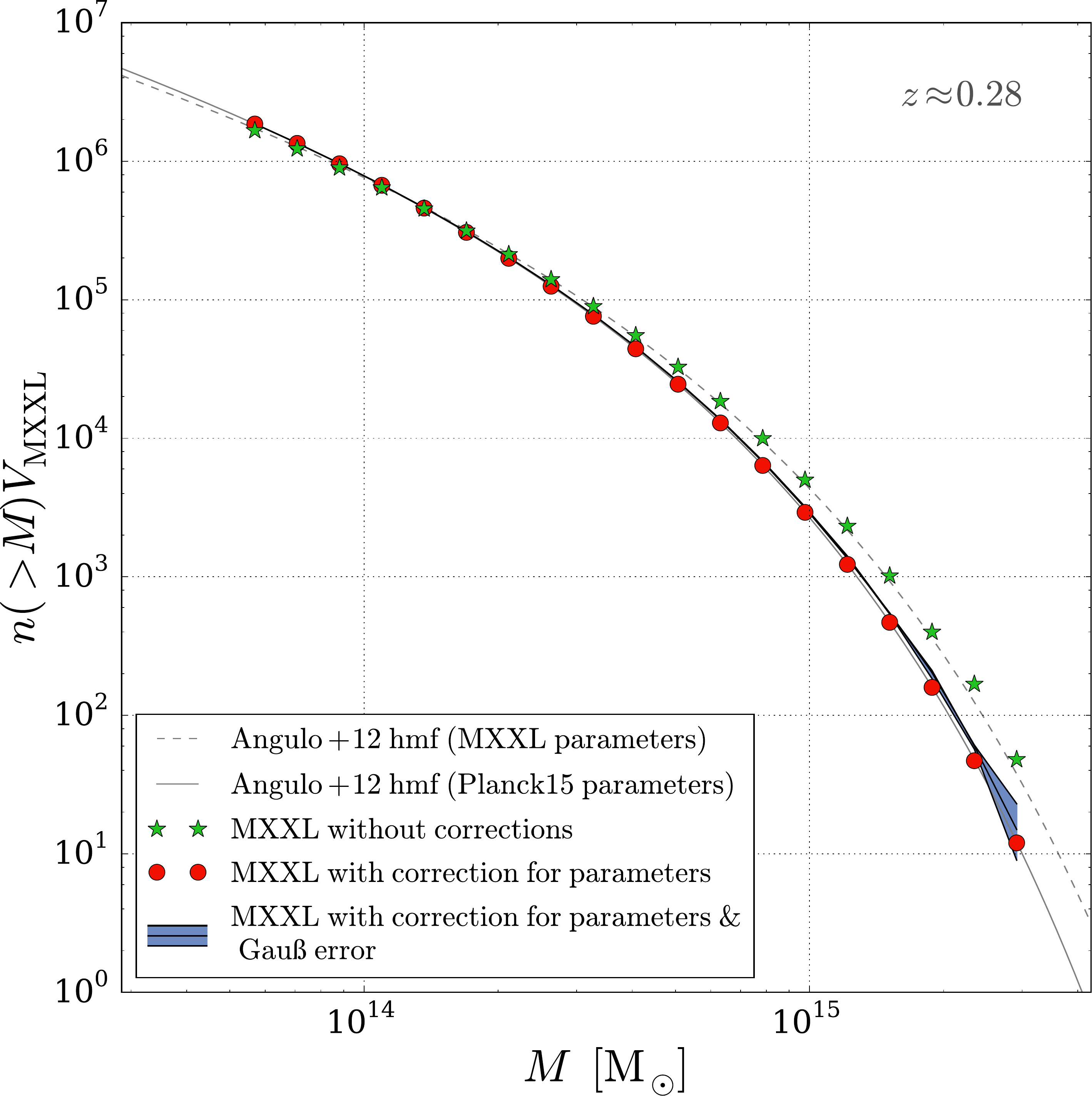}
    \caption{The cumulative subhalo mass function after applying the corrections discussed in Section~3.3. The lines show the halo mass functions for subhaloes with the original MXXL cosmological parameters (dashed) and the {\it Planck} parameters (solid), using the formulae from \citet{Angulo2012}. The symbols show the cumulative mass function of MXXL subhaloes without any corrections (green stars) and after correcting the mass distribution to the updated {\it Planck} cosmological parameters (red dots). The range of different realisations of the correction for mass errors from the lensing analysis is shown by the blue band.}
    \label{fig:hmf}
\end{figure}

The impact of the choice of cosmological parameters on the halo mass function at $z = 0.28 $ is shown by the lines in Fig.~\ref{fig:hmf}. The decrease in $\sigma_8$ in the {\it Planck} cosmology compared to the value used in the MXXL leads to a drop in number of haloes at the high mass end of the mass function. However, the higher value of $\Omega_{\rm m}$ in the Planck cosmology has the opposite effect and partially compensates for the change in $\sigma_{8}$. The net effect is a reduction in the abundance of haloes above a mass of $3 \times 10^{14} \, \Msun$.

To correct the halo masses for the change in the cosmological parameters, we first rank all FoF haloes by mass down to a mass of $10^{13}\,\Msun$, which is well below the masses of interest for the comparison with the substructures of Abell 2744. We then assign each halo a new mass according to the mass this rank is expected to have with the {\it Planck} cosmological parameters. The same procedure is applied to all subhaloes with a mass exceeding $10^{13}\, \Msun$. Fig.~\ref{fig:hmf} shows that the corrected masses follow the halo mass function corresponding to {\it Planck} cosmological parameters. We note that the halo mass function of FoF haloes in MXXL differs significantly from that of subhaloes \citep{Angulo2012}. We therefore use different fitting formulae for FoF haloes and subhaloes, as presented in \citet{Angulo2012} (Eqs.\,2 and 3 which were obtained by fitting to all three Millennium simulations). Halo mass functions are calculated with the Python module {\tt hmf} \citep[][]{Murray2013} which contains implementations of both mass function fits from \citet{Angulo2012}.

The use of different cosmological parameters not only affects the shape of the halo mass function, but can also change the halo merger rate. The merger rate can be estimated using the merger probability presented in \citet[][]{Lacey1993}. Adapting their discussion, the instantaneous merger probability, i.e. the probability that a halo of mass $M_1$ accretes a mass $\Delta M$ to form a halo of mass $M_2 = M_1 + \Delta M$ within a scale factor change of $\mathrm{d}\ln a$ is  
\begin{equation}
\begin{aligned}
\frac{\mathrm{d}^2 p}{\mathrm{d}\!\ln\Delta M\,\mathrm{d}\!\ln a} =  &\left(\frac{2}{\pi}\right)^{1/2}\left|\frac{\mathrm{d} \ln \delta_c}{\mathrm{d} \ln a}\right| \left(\frac{\Delta M}{M_2}\right)\\
&\times \left|\frac{\mathrm{d} \ln \sigma_2}{\mathrm{d} \ln M_2}\right|\frac{\delta_c(a)}{\sigma_2} \frac{1}{(1-\sigma_2^2/\sigma_1^2)^{3/2}}\\
& \times\exp\left[-\frac{\delta_c(a)^2}{2}\left(\frac{1}{\sigma_2^2}-\frac{1}{\sigma_1^2}\right)\right]\ ,
\end{aligned}
\end{equation}
where $\sigma_1 \equiv \sigma(M_1)$ and $\sigma_2 \equiv \sigma(M_2)$ denote the variance of the density contrast after smoothing with a window function containing mass $M_1$ or $M_2$, respectively, and $\delta_c(a)$ is the critical density contrast at scale factor $a$ at which a region collapses according to linear theory. Changing the cosmological parameters leads to an increase of $9 \%$ in the instantaneous merger probability for two haloes of masses $M_1 = \Delta M = 2\times10^{14}\,{\rm M}_{\odot}$ at $z = 0.308$, from a value of $0.278$ with the MXXL cosmological parameters to a value of $0.303$ with the {\it Planck} parameters. The merger rate at the time scale and mass range relevant to Abell 2744 is therefore underpredicted in the MXXL compared to a model with {\it Planck} parameters. However, it is important to bear in mind that the merger probability does not increase on all time scales. When integrated over time $\mathrm{d}p/\mathrm{d}\!\ln \Delta M$, the merger rate decreases by 18\,\% from $0.216$ to $0.178$ on using the updated cosmological parameters. Hence, the cosmological parameters adopted in the MXXL could have a significant influence on the positions and numbers of subhaloes. It is, however, difficult to correct for these effects without re-running the simulation with updated parameters. Since a higher merger rate is needed to find objects similar to Abell 2744, the enhanced instantaneous merger rate is the only effect which would help to increase the probability of finding an Abell 2744 like cluster in the MXXL simulation.

\subsection{Correction for Eddington Bias: including mass errors}
The  error in the observational mass measurements causes a spurious increase in the number of high mass objects, which is known as Eddington bias \citep{Eddington1913}. The statistical error introduced by the lensing analysis leads to a random perturbation in mass around the true value. In the regime where the mass function declines steeply, i.e. the regime relevant for cluster-mass dark matter haloes, it is statistically more probable that a lower mass halo will scatter up in mass than vice versa. Hence the errors result in more high mass haloes compared to the true, underlying mass function and produce a bias in the observed halo mass function which depends on the precision of the mass determination.

We can incorporate this effect into our comparison by drawing a new halo mass from a Gaussian distribution whose width is set by the estimated error on the observed mass. The mean of this Gaussian is given for each halo by the mass after the correction with respect to updated cosmological parameters. The errors for most halo masses are in the range of $7.5-12$\,\% (see Table~\ref{tab:substructures}). Only the {\it S4} subhalo mass has a relative error of 22\,\%. Thus, we chose an error of 15\,\% for the Gaussian. In contrast, the error of the {\it Core} subhalo mass is more than one order of magnitude smaller. For this subhalo, weak and strong lensing data were used to reconstruct the mass potential which results in a considerably more precise mass determination. We repeated the random sampling of masses 100 times for subhaloes and 500 times for FoF haloes to estimate the uncertainty introduced by this procedure. We show three examples of the 100 subhalo realisations in Fig.~\ref{fig:hmf} indicated by the blue band; these correspond to the two realisations with the highest and lowest number of haloes with mass $M = 2.9 \times 10^{15} \, \Msun$ and the one with the median number of such haloes. Accounting for the Eddington bias leads, as expected, to a modest increase in the mass function at the high mass end ($M > 2 \times 10^{15} \,\Msun$). At masses lower than this there is no discernible change in mass function.

\section{Searching for Abell 2744 in the MXXL simulation}
\label{sec:searchMXXL}
We are now in a position to estimate the chance of finding a cluster like Abell 2744 in $\Lambda$CDM. We carry out two comparisons with the adjusted MXXL haloes: (1) the total mass of Abell 2744 and (2) the masses and distances of the substructures from the cluster centre.

After applying the mass extrapolation and corrections described in Sections~\ref{sec:massSimuObs} and \ref{sec:massMXXLCorr}, we searched for FoF haloes with mass within the {$3\sigma$-range} of the aperture mass of Abell 2744 ($M(R<1.3\,{\rm Mpc}) = \left( 2.3 \pm 0.1 \right) \times 10^{15}\,{\rm M}_\odot$) from \cite{Jauzac2016}. In contrast to the analysis in \cite{Jauzac2016} we consider 500 different random realisations when perturbing the FoF masses to account for Eddington bias. Furthermore, we applied our search to the extrapolated $M_{200}$ masses and not to extrapolated $M_{\rm FoF}$ masses. We find on average $68 \pm 6$ haloes which meet the Abell 2744 mass condition at $z = 0.32$ and $92 \pm 7$ at $z = 0.28$. The maximum number of haloes meeting the mass criterion was 86 haloes at $z = 0.32$ and 113 haloes at $z = 0.28$. These numbers are higher than the 39 clusters reported by \cite{Jauzac2016} where we only considered a single random realisation of the Eddington bias and used $M_{\rm FoF}$ masses. However, the qualitative result that haloes as massive as Abell 2744 are quite common in the MXXL simulation remains unchanged.

To find MXXL haloes with comparable substructure to Abell 2744, we analysed the properties of subhaloes identified by {\tt SUBFIND}. We put a rectangular box around each subhalo and counted the number of subhaloes with suitable mass within this box. To account for projection effects, we took the box depth ($ 15 \, {\rm Mpc}$) to be larger than the length of the end face sides ($2 \,{\rm Mpc}$) and only consider the projected 2D distances. The value of $15\,{\rm Mpc}$ was chosen as it is comparable to the displacements introduced by peculiar velocities for an object as massive as Abell 2744. The search was performed using the $x$-, $y$- and $z$-axis in turn as the line-of-sight. To complete the search efficiently we only considered subhaloes with mass above $5 \times 10^{13} \, {\rm M}_{\odot}$.

We found no halo in the MXXL simulation with eight subhaloes as massive and as close to the halo centre as found in Abell 2744 (see Table~\ref{tab:extrapolated_masses}). Even if we increase the box depth to $30\,{\rm Mpc}$ we find no suitable haloes. Larger projection depths are unlikely as they would have given rise to a separate peak in the observed galaxy redshift distribution. 
\begin{figure}
	\includegraphics[width=\columnwidth]{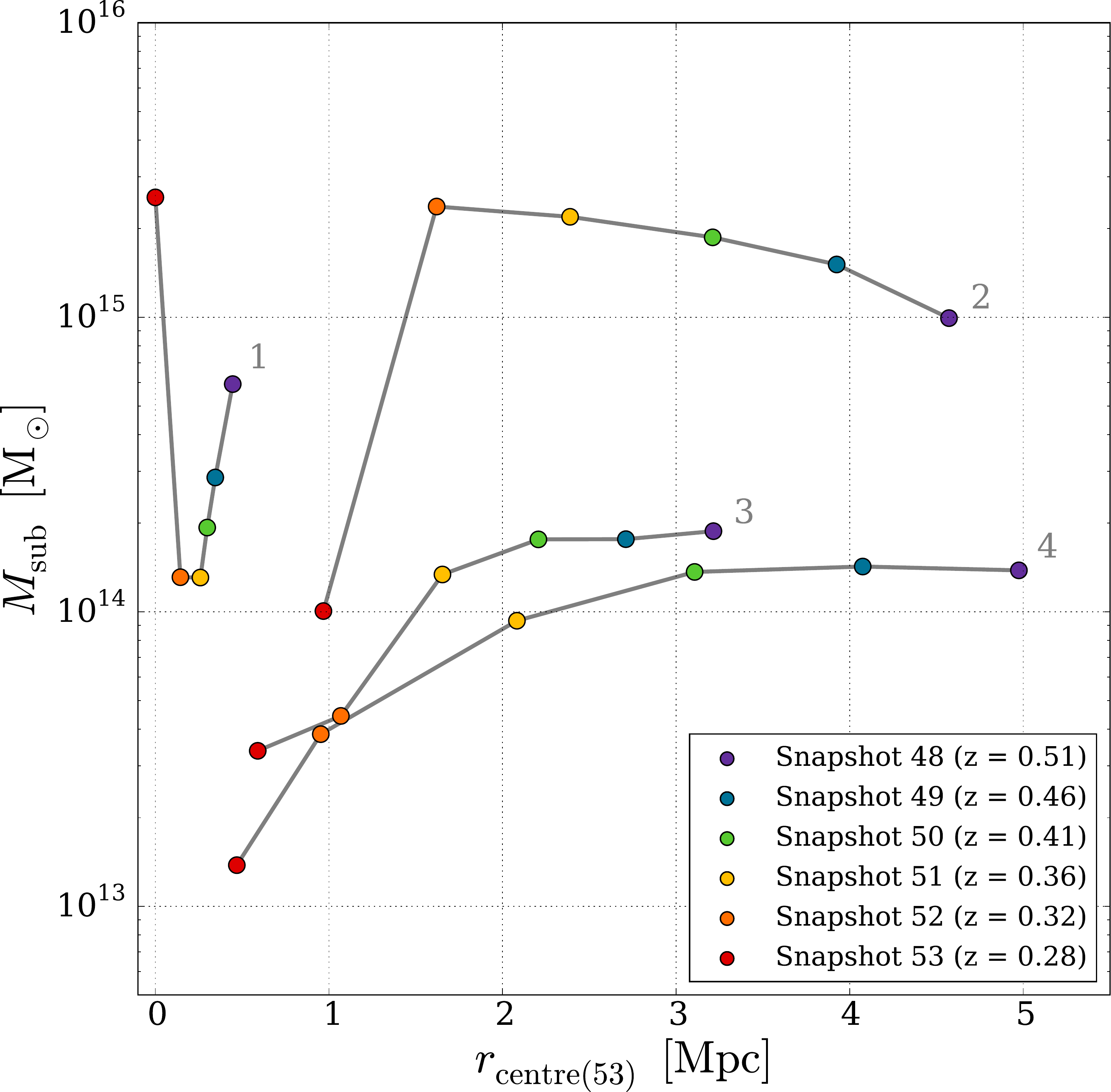}
    \caption{Time evolution of the mass and radial distance of the four most massive subhaloes in an MXXL halo with similar mass to Abell 2744 ($M_{200} = 3.1 \times 10^{15} \, {\rm M}_\odot$). The six different snapshots (48-53) are coded by colours as given in the legend. The radius is the distance from the position of the central halo at the final snapshot (number 53).}
    \label{fig:substructurePaths}
\end{figure}%

We can estimate the chance that substructures are projected over a much larger distance than 30\,Mpc by using the Press-Schechter formalism \citep{Press1974}. We calculate the number of haloes predicted by the Press-Schechter mass function in the patch of the sky observed in \cite{Jauzac2016}. We consider projections over a redshift range between z=0.2 and 0.8, since gravitational lensing is most efficient when the lens is midway between source and observer. This results in 0.3 haloes with a mass $M \ge 5\times 10^{13}\,\Msun$  being projected in a 2 by 2\,Mpc patch at z=0.3. Thus, it seems very improbable that more than one of the subhaloes is a line-of-sight projection unrelated to the galaxy cluster.

Our failure to find a halo in the MXXL with substructure comparable to Abell 2744 is in agreement with the results of \cite{Munari2016}. These authors determined the velocity dispersion of member galaxies of Abell 2142 and used this as a proxy for the substructure masses. The observations were compared to subhaloes of comparable virial mass in numerical simulations of different resolution which included baryonic physics. They found a smaller number of massive (in their case, circular velocity subhaloes above 200 ${\rm km s}^{-1}$) with a significance level of 7$\sigma$. This shows that the problem is not caused by subtleties of the gravitational lensing analysis or differences between mass estimates.

\section{Tidal stripping}
\label{sec:tidalStripping}

As pointed out in \cite{Jauzac2016}, dynamical effects such as tidal stripping can have a significant influence on the identification and masses of subhaloes in N-body simulations. The properties of subhaloes are also sensitive to the manner in which they are found. \cite{Muldrew2011} demonstrated that tidal stripping of infalling subhaloes is overpredicted by the {\tt SUBFIND} algorithm. This is caused by the algorithm identifying substructures through the presence of saddle points in the density profile. The closer infalling subhaloes get to the central halo the higher the density becomes and the harder it is for {\tt SUBFIND} to identify all of the particles which belong to the subhalo. All eight substructures in Abell 2744 lie within a radius of $1 \, {\rm Mpc}$ from the core which is less than half the virial radius ($R_{200} \approx 2.8 \, {\rm Mpc}$). The subhalo masses could be underpredicted by more than 50\% according to \cite{Muldrew2011}. However, a comparison with other subhalo finders in \cite{Behroozi2015} showed that {\tt SUBFIND} is still one of the most reliable available substructure finders. Nevertheless, it is still important to take this possible bias into consideration, especially when looking close to the cluster centre. 

We investigate the effect of tidal stripping for several haloes with a similar mass to Abell 2744 in the MXXL. In Fig.~\ref{fig:substructurePaths}, we show the evolution of four subhaloes in a halo with mass $M_{200} = 3.1 \times 10^{15} \, {\rm M}_\odot$ (after applying the corrections described in Section~\ref{sec:massMXXLCorr}). Halos 3 and 4 lose roughly one order of magnitude in mass as they approach the halo centre. A loss of mass due to tidal stripping is indeed expected in this situation, but as pointed out in \cite{Muldrew2011}, the {\tt SUBFIND} algorithm overpredicts the amount of stripping. Fig.~\ref{fig:substructurePaths} also shows an interesting apparent exchange of mass between haloes 1 and 2 between snapshot 52 and 53. This could be due to the treatment of dark matter particles that are gravitationally bound to the whole FoF halo rather than to one of the identified subhaloes. {\tt SUBFIND} assigns these particles to the most massive subhalo. If the most massive subhalo changes from one snapshot to another and a large amount of dark matter is not directly bound to a subhalo, it would cause a shift in mass between subhaloes.

\begin{figure*}
\begin{center}
\vspace{2em}
\includegraphics[width=1.85\columnwidth]{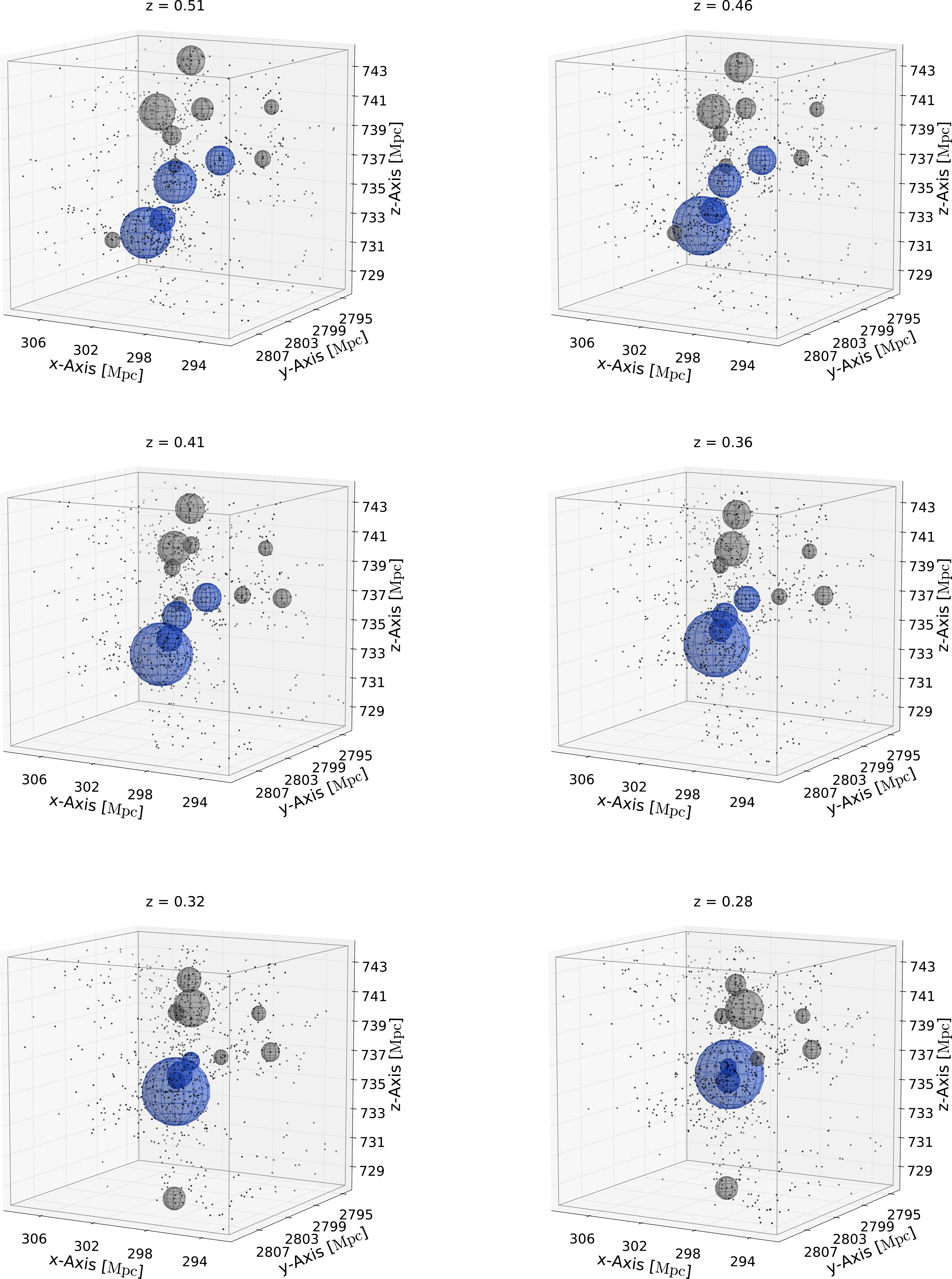}
\vspace{2em}
\caption{\label{fig:stripping3d}Time evolution of the environment of the FoF halo shown in Fig\,\ref{fig:substructurePaths}. Each panel shows a snapshot in the MXXL from number 48 to 53 with the corresponding redshifts labelled above each panel. Subhaloes above $10^{13} {\rm M}_{\odot}$ are shown as spheres with radius scaled by their $R_{200}$. The four subhaloes shown in Fig.~\ref{fig:substructurePaths} are highlighted as blue spheres. All other subhaloes are shown as dots.}
\end{center}
\end{figure*}

We show the same cluster and the time evolution of its environment in Fig.~\ref{fig:stripping3d}. Here the radius of the sphere plotted is set according to $R_{200}$. Subhaloes with mass above $10^{13} \, {\rm M}_{\odot}$ are shown. The four subclusters plotted in Fig.~\ref{fig:substructurePaths} are highlighted as blue spheres in Fig.~\ref{fig:stripping3d}. All other subhaloes with mass below  $10^{13} \, {\rm M}_{\odot}$ are plotted as dots. The time sequence shows that the cluster is undergoing a complex merger which involves more than ten massive subhaloes. As in Fig.~\ref{fig:stripping3d}, the mass loss of the infalling subhaloes due to tidal stripping can be seen. The radii of the infalling haloes (except for the central halo) shrink considerably from snapshot 48 ($z = 0.51$) to 53 ($z = 0.28$). When considering the last two snapshots 52 ($z = 0.32$) and 53 ($z = 0.28$), the environment reveals no explanation for the jump in mass between haloes 1 and 2. This plot shows that both the overprediction of tidal stripping as well as the assignment of mass which is not bound to any subhalo but to the whole system can influence the masses of subhaloes significantly.

\begin{figure*}
\centering
\begin{tabular}{cc}
	\subfloat{\includegraphics[height=0.75\columnwidth]{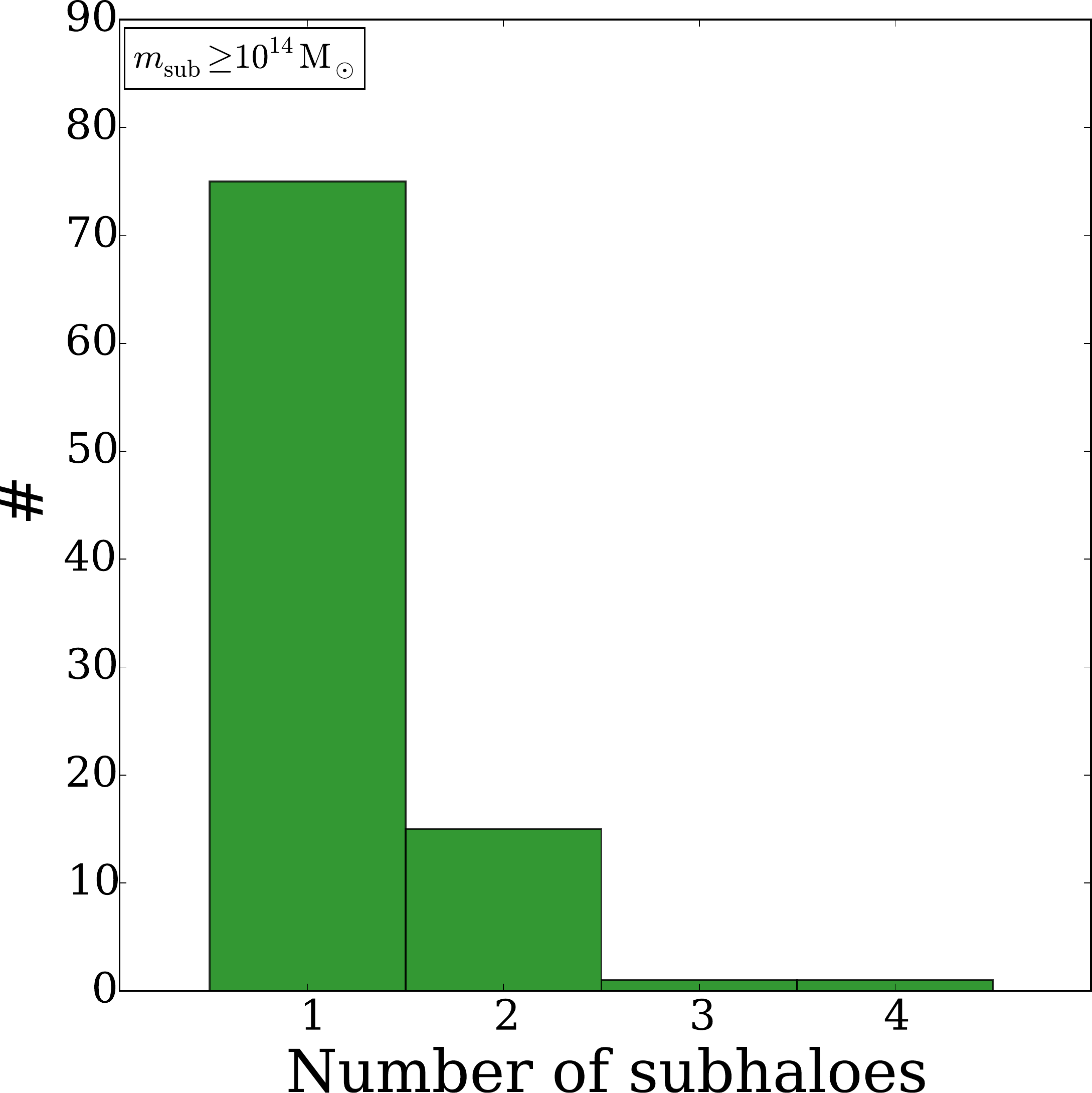}
	\label{subfig:distribution1}} &
	\subfloat{\hspace{2em}\includegraphics[height=0.75\columnwidth]{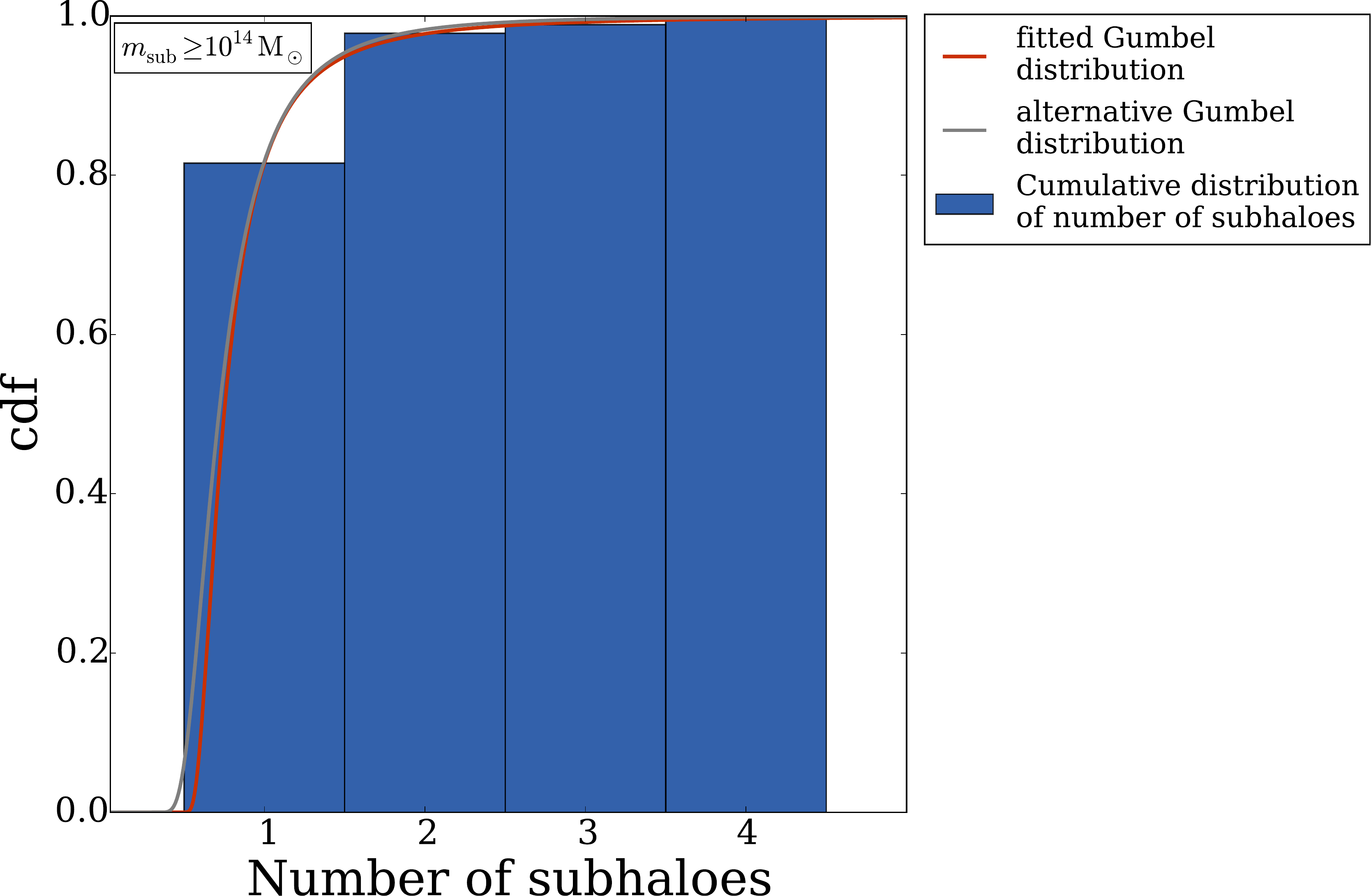}
	\label{subfig:distribution2}} \\
	\subfloat{\includegraphics[height=0.75\columnwidth]{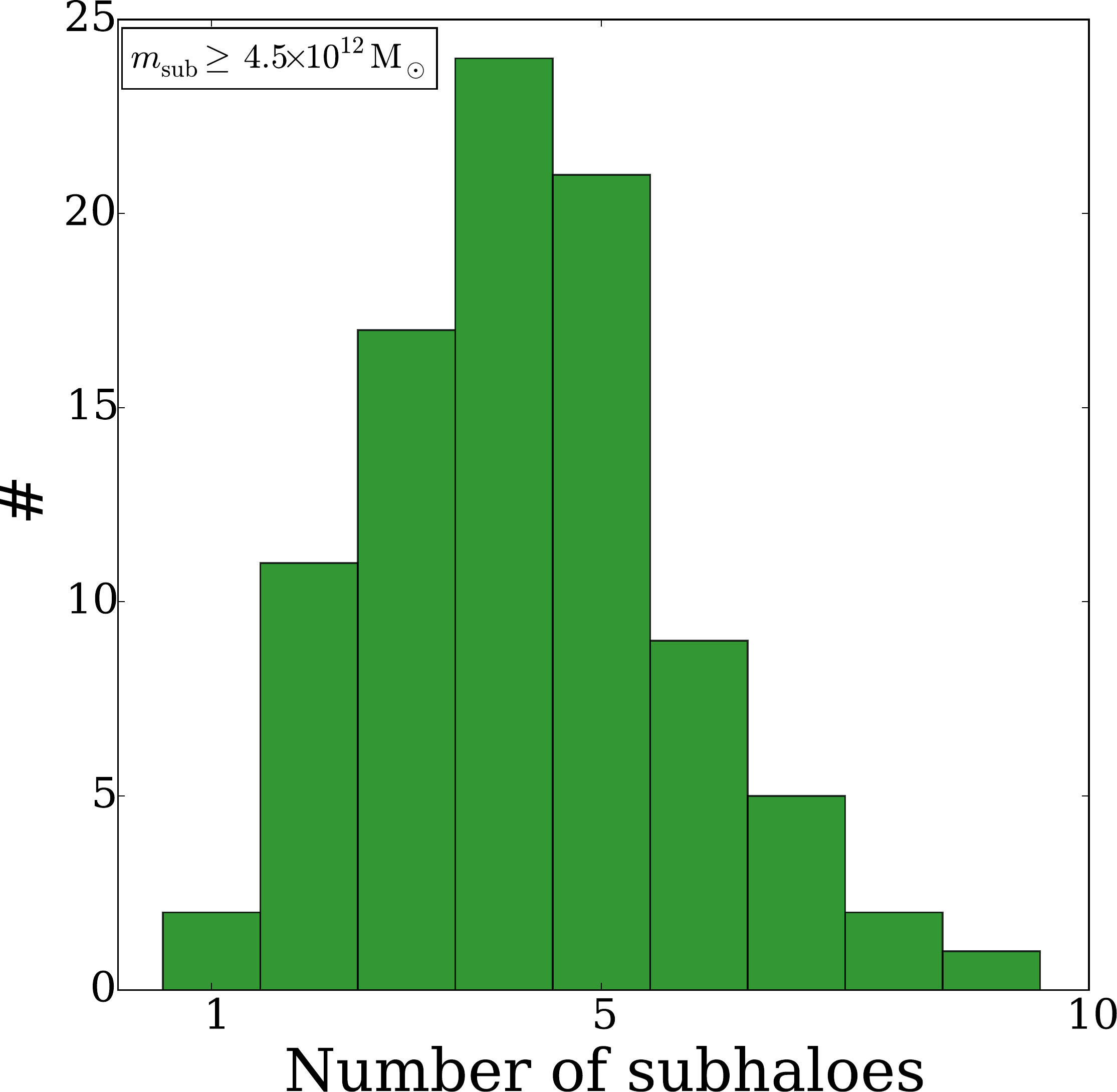}
	\label{subfig:distribution1}} &
	\subfloat{\hspace{2em}\includegraphics[height=.75\columnwidth]{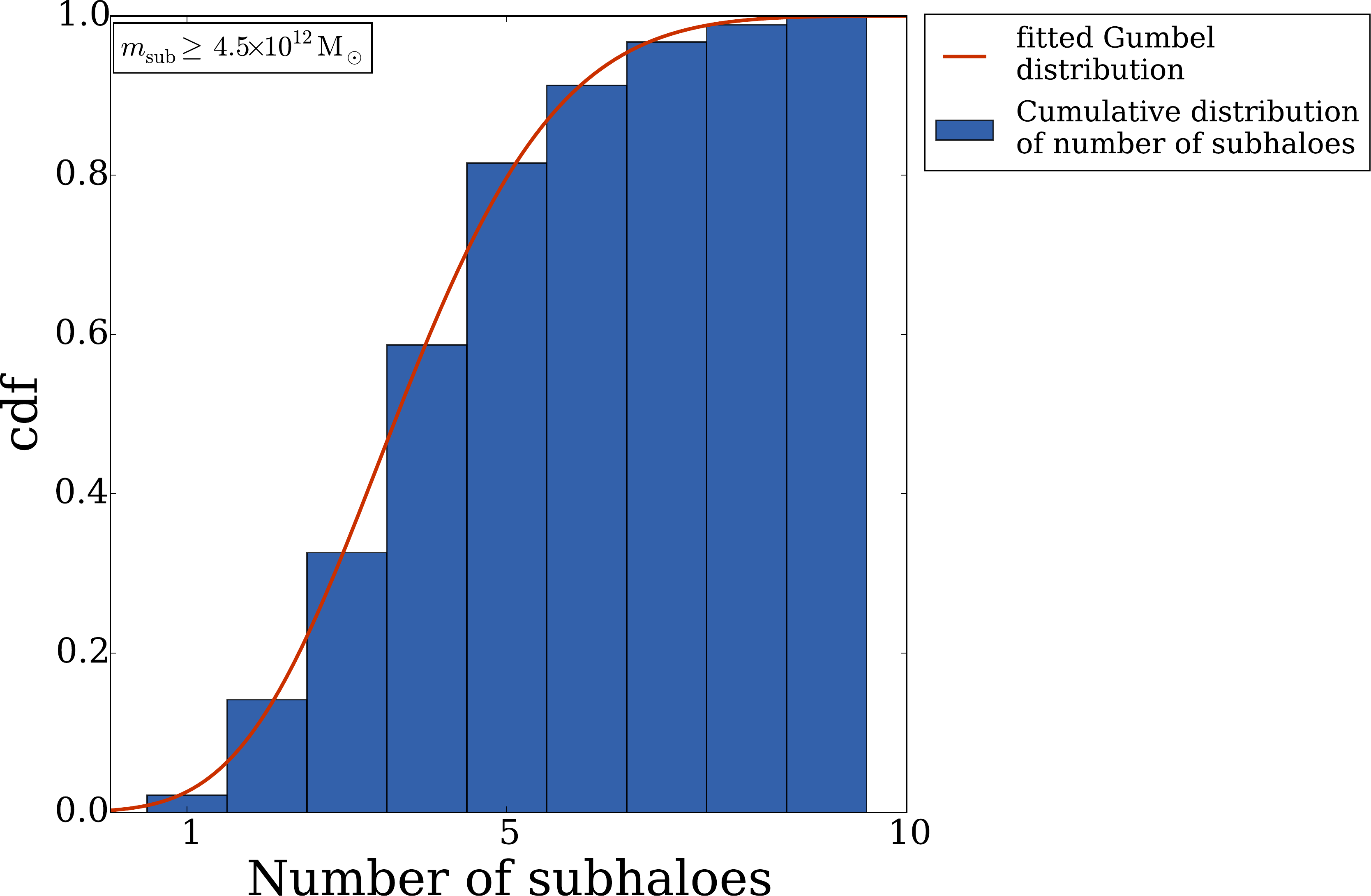}
	\label{subfig:distribution2}} \\
\end{tabular}
\caption{\label{fig:distribution}The distribution of the number of massive substructures in Abell 2744-like haloes in the MXXL. The upper panels show distributions for subhaloes with masses $m_i \geq 10^{14} \, \Msun$, whereas the lower panels show subhaloes above a lower mass threshold ($m_i \geq 4.5 \times 10^{12} \, \Msun$). The left panels show the distribution of the number of subhaloes with mass above the threshold, the right panels show the cumulative distribution of subhalo numbers with a fitted extreme value distribution (red line), and in the top right panel an extreme value distribution with different parameters (grey line) to emphasize the low significance for the $m_i \geq 10^{14} M_\odot$ case. The parameters of all three extreme value distributions are listed in Table~\ref{tab:fittingParameters}.}
\end{figure*}%

We searched for clusters with Abell 2744-like substructure while trying to reduce the impact of tidal stripping to a minimum. We did this by taking all clusters with a similar total mass to Abell 2744. We traced the masses of all subhaloes that end up within a radius of $1\,{\rm Mpc}$ from the centre back in time. We then found the snapshot where the number of subhaloes above a threshold of $10^{14}\,\Msun$ reached its maximum. The subhalo mass at this snapshot, i.e. the subhalo mass before infall, was used instead of the mass after infall. Despite the fact that this procedure completely neglects any tidal stripping, we still did not find any clusters with a similar distribution of substructure to that in Abell 2744.

We found one cluster with a mass compatible with Abell 2744 that has 4 subhaloes above $10^{14} \, \Msun$ and another cluster with 3 such subhaloes. Roughly 15\% of the haloes with the right mass contain two subhaloes and the majority (more than 80\%) have only one subhalo above the threshold. Thus, tidal stripping alone cannot explain the discrepancy between observations and the MXXL simulation. Artificially minimising the effects of tidal stripping, however, increases the number of subhaloes above $10^{14}\,\Msun$ in one case by a factor of 3 and in $\sim 18\%$ of the cases by at least a factor of 2.

\section{Extreme value statistics}
\label{sec:extremeVal}

Extreme value statistics allow us to assess the probability of finding unusual objects in the Universe \citep[e.g.][]{Davis2011,Waizmann2011,Yaryura2011,Reischke2015}. Since we did not find a halo in the MXXL simulation with a distribution of substructure like that seen in Abell 2744, we can use extreme value theory to estimate the simulation volume that would be needed to find at least one halo with the same substructure. 

It was shown by \cite{Mises1936} and \cite{Jenkinson1955} that the block maximum $M_n = \max \left( X_1,...,X_n\right)$ of a set of random variables $\{X_i\}$ follows a cumulative density function described by
\begin{equation}
\label{eq:Gumbel}
G_{\alpha,\beta,\gamma}(x) = \left\{
\begin{array}{ll}
\exp\left(-\left[1+\gamma\left(\frac{x-\alpha}{\beta}\right)\right]^{-1/\gamma}\right),& \textrm{for } \gamma\neq 0 \\[1.5em]
\exp\left(-\exp\left(-\frac{x-\alpha}{\beta}\right)\right),& \textrm{for } \gamma= 0\, .
\end{array}
\right.
\end{equation}%
We use this expression to predict the probability of finding a cluster with a similar number of substructure to Abell 2744.

We apply extreme value statistics to the number of substructures in haloes found after applying the corrections described in Section~\ref{sec:massMXXLCorr}. After updating the cosmological parameters, we corrected for the Eddington bias, making 500 realisations to include the statistical uncertainty. Here we show one of these realisations in which 92 FoF haloes with a mass consistent with Abell 2744 were found at $z = 0.28$. We also minimise tidal stripping by using the masses before the infall as described in Section~\ref{sec:tidalStripping}.  Fig.~\ref{fig:distribution} shows histograms of the number of subhaloes within $1 \,{\rm Mpc}$ of the main halo centre with mass above thresholds of $4.5 \times 10^{12} \, \Msun$ and $10^{14} \, \Msun$. The lower value of the threshold ($4.5 \times 10^{12} \, \Msun$) was chosen to expect one halo with at least eight subhaloes above the mass threshold in the MXXL simulation.

We fit an extreme value distribution (Eq.~\ref{eq:Gumbel}) to the cumulative density function. The best-fitting parameters for the extreme value distributions are listed in Table~\ref{tab:fittingParameters}. In the case of the lower mass threshold, the cumulative distribution function is well described by the extreme value distribution. We use a Kolmogorov–Smirnov test to quantify the agreement of the theoretical distribution with the one derived from the MXXL simulation. The test statistic $D_n$ is given by
\begin{equation}
D_n =  \sup_{x_i}|F_n(x_i) - F(x)|,
\end{equation}
where $n$ is the number of samples (here $n = 92$), the $x_i$ are the observed values and $F(x)$ is the theoretical cumulative distribution function. In the case of the lower mass threshold we find $D_{92} = 0.018$. The corresponding $p$-value (i.e. the probability of obtaining this test statistic or a more extreme one) can be calculated as $p \approx 1$. The high $p$-value shows that the statistic is indeed well described by the extreme value distribution. However, it should be pointed out that the $p$-value is slightly overpredicted, since the distribution was fitted to the data. The good agreement between the distribution obtained from the simulation and the fit can also be seen in the lower right panel of Fig.~\ref{fig:distribution}.

In the case of the higher mass threshold of $m_i \geq 10^{14} \, \Msun$ the maximum number of subhaloes found is four. Nevertheless, we use the fitted statistic to get a rough estimate of the probability of finding a halo in the MXXL with a substructure distribution like that of Abell 2744. The fitted GEV statistic gives the probability of finding the required amount of substructure as $P(N\geq8|M_{\rm FoF} = M_{\rm A2744};m_i\geq 10^{14} \, \Msun) = 0.073\%$. Since we find 92 Abell like clusters in the MXXL simulation, we would need a simulation with a volume $\sim 14$ times larger as MXXL. This result is rather speculative. A slight change to the distribution (as shown by the grey line in Fig.~\ref{fig:distribution}) decreases the probability by more than a factor of four to $P = 0.017 \%$. With this the volume would have to be $\sim 57$ times larger than that of the MXXL and thus more than 570 times the volume of the observable Universe up to $z=0.308$. In both cases we would need a significant increase in the volume of the simulation, which shows that the observation of Abell 2744 is in tension with the predictions of MXXL even with the corrections applied. This is clear from the fact that we have to reduce the mass threshold for subhaloes to $4.5\times10^{12}\,\Msun$ to find an example with eight substructures close to the centre. This threshold is two orders of magnitude less massive than the observed substructures, which suggests a deeper problem than simple bad luck. 

\begin{table}
  \caption{Fitted parameters of the three extreme value distributions (Eq. \ref{eq:Gumbel}) shown in Fig.~\ref{fig:distribution}.}
  \centering
\resizebox{0.48\textwidth}{!}{%
    \begin{tabular}{lccc}
  \hline\hline
  \noalign{\smallskip}
     & $\alpha$ & $\beta$ & $\gamma$ \\[0.2em] \hline 
\noalign{\smallskip}$m \geq 4.5\times10^{12}\,\Msun$ & $3.120 \pm 0.001$ & $1.448 \pm 0.002$ & $-0.185 \pm 0.001$\\
\noalign{\smallskip}$m \geq 10^{14}\,\Msun$ (fit)& $0.696 \pm 0.037$ & $0.131 \pm 0.014$ & $0.446 \pm 0.085$\\
\noalign{\smallskip}$m \geq 10^{14}\,\Msun$ (alt.)& 0.65 & 0.17 & 0.30\\
    \hline\hline
  \end{tabular}
  }
\label{tab:fittingParameters}
\end{table}%

\section{Summary}
\label{sec:summary}
We investigated various effects that can influence the comparison of masses estimated from gravitational lensing analyses and from simulations. 
On the observational side the lensing mass estimate is affected by:
\begin{itemize}[leftmargin=2em]
\setlength{\itemindent}{-0.5em}
\item the definition of the mass estimate (i.e. comparing the projected mass enclosed within a circular aperture of radius $1.3 \, {\rm Mpc}$ with the mass enclosed within a sphere of 
radius $R_{200}$)
\item line-of-sight projections\vspace{-0.5em}
\begin{itemize}
\setlength{\itemindent}{-0.5em}
\item the adding up of mass along the line-of-sight within the aperture
\item the erroneous assignment of haloes behind the cluster or between cluster and observer as substructure of the cluster\vspace{-0.5em}
\end{itemize}
\item the Eddington bias due to errors in the inferred mass and the steepness of the halo mass function
\end{itemize}
On the simulation side several effects can bias the mass obtained for substructures, such as:
\begin{itemize}[leftmargin=2em]
\setlength{\itemindent}{-0.5em}
\item the mass definition (e.g. $M_{200}$, $M_{\rm FoF}$ or $M_{\rm sub}$)
\item the choice of cosmological parameters which \vspace{-0.5em}
\begin{itemize}
\item affects the halo mass function
\item affects the merger probability \vspace{-0.5em}
\end{itemize}
\item overpredicted tidal stripping and how subhaloes are found by the substructure finding algorithm
\item the assignment of mass belonging to the whole system to the most massive subhalo
\end{itemize}

We found that the $M_{200}$ mass of Abell 2744-like haloes in MXXL is 40\% larger than the mass estimated within the aperture of $1.3 \, \rm{Mpc}$ used in the gravitational lensing analysis. For subhaloes this effect is even more important as in this case the radius of the cylinder is significantly smaller than $R_{200}$. The extrapolated {\tt SUBFIND} masses are all an order of magnitude larger than the estimates obtained from the gravitational lensing analysis. This is due to the relatively small aperture used of $150 \, \rm{kpc}$, which was chosen to minimise overlap of close subhaloes.

Line-of-sight projections were taken into account in three ways. At first we integrated over a cylinder instead of a sphere to match the mass estimates. Furthermore, we projected all haloes within a box with a depth of $15\,{\rm Mpc}$ onto a 2D map. We then only considered 2D distances. We find that a projection of haloes over a larger distance on the sky is rather improbable. The Press-Schechter formalism predicts that only 0.3 haloes with a mass of $M>5\times10^{13}\Msun$ are projected onto the observed patch of the sky. We also estimated the effect of combining subhaloes along the line-of-sight by adding up the masses of all subhaloes within a cylinder with a radius of $150 \, {\rm kpc}$ and depth of $15 \, {\rm Mpc}$. This led to the result that maximally one subhalo is scattered above $10^{14} \, {\rm M}_{\odot}$. Therefore, projection effects could also be responsible in part for the observation of many massive substructures in Abell 2744.

We corrected for the choice of cosmological parameters used in the MXXL simulation. We assigned each halo and subhalo a new mass according to the halo mass function expected in the {\it Planck} cosmology. The number of haloes above $10^{15} \, \Msun$ decreased to roughly two thirds of the original abundance. The change of merger probability and different particle coordinates, however, was not considered. These effects are hard to correct without re-running the simulation. 

The Eddington bias was taken into account by sampling random masses for each halo. The masses were drawn from a Gaussian distribution with a mean corresponding to the mass after updating the cosmological parameters and with a standard deviation of the estimated lensing error. We repeated this process 100 times for subhaloes and 500 times for FoF haloes to estimate the statistics of the outcome. This correction had a small but still noticeable effect, especially at the very high mass end of the halo mass function. The number of haloes with a mass above $10^{15} \, \Msun$ was on average increased by 10\% and maximally by 13\% after including the mass errors. 

The feature of {\tt SUBFIND} of assigning all particles that are not bound to a particular substructure to the most massive subhalo was also partly taken into account. We considered this by not applying any upper mass limit to the central halo. However, all smoothly distributed dark matter between the subhaloes which adds to the observed substructure masses cannot be added to the subhaloes in MXXL retroactively.

After applying these corrections to MXXL haloes, we find on average $68 \pm 6$ haloes (the most extreme example we find contains 86 halos) at redshift $z = 0.32$ with a mass similar to Abell 2744. At redshift $z = 0.28$, we find on average $92 \pm 7$ haloes (with an extreme example with 113 halos). We do not find any halo with comparable substructure to Abell 2744 in the MXXL. The maximum number of massive subhaloes within a radius of $1 \, {\rm Mpc}$ was 3, much less than the observed number of eight massive substructures in Abell 2744. This could be explained in part by the fact that not all effects could be taken into account (i.e.\ the different merger rate, mass assignment of {\tt SUBFIND}). It is also possible that the time resolution given by the two snapshots at $z = 0.32$ and $z = 0.28$ is insufficient to capture the complex dynamical state of Abell 2744.

Furthermore, {\tt SUBFIND} possibly overpredicts the tidal stripping of subhaloes. To investigate this effect for the haloes with a total mass comparable to Abell 2744, we traced the masses of each subhalo back in time and looked for the snapshot with the most subhaloes above a mass threshold of $10^{14} \, \Msun$. In one case, this increased the number of subhaloes above $10^{14} \, \Msun$ from 1 to 3 and in another case from 2 to 4. We conclude that individually none of the effects we have investigated come close to reconciling the simulation results with the observed substructure distribution in Abell 2744. 

We used extreme value statistics to estimate how large a simulation volume would be needed to find a halo like Abell 2744. The simulation volume would have to be $\sim 14$ times larger than MXXL. However, the MXXL simulation already corresponds to ten times the volume out to $z~=~0.308$ in which Abell 2744 is found. To find one cluster with eight substructures in MXXL, the mass threshold has to be lowered dramatically to a value of $4.5 \times 10^{12} \, \Msun$. This threshold is two orders of magnitude lower than the mass of the observed subclusters. This proves that it is not only a matter of being ``unlucky'' with the random haloes in the simulation, but hints that there is a deeper problem either with the compared mass estimates or with $\Lambda$CDM itself. 

To relieve this tension within the $\Lambda$CDM model, higher values of $\Omega_m$ and $\sigma_8$ would be needed to achieve a higher number of objects of suitable mass (see Fig.\,\ref{fig:hmf}). However, the values of $\Omega_m$ and $\sigma_8$ from the \emph{Planck 2015} set of cosmological parameters are already high in comparison to the values obtained from gravitational lensing \citep[e.g.][]{Joudaki2016, Hildebrandt2016}. Apart from $\Lambda$CDM, \cite{Jauzac2016} discussed how the result would change if a different type of dark matter is considered. They took into account a warm dark matter (WDM) or a self interacting dark matter (SIDM) particle. They find that the abundance of subhaloes more massive than $10^9 \Msun$ in WDM is nearly identical to that of CDM. Since the substructures of Abell 2744 considered here are more massive than $10^{14} \Msun$, the discrepancy is insensitive to the choice between WDM and CDM.  When analysing the cross section of a possible SIDM particle, \cite{Jauzac2016} find that SIDM is not favoured over CDM by the substructure distribution observed in Abell 2744.

Previous work comparing the abundance and distribution of substructures has tended to focus on a somewhat lower mass range than we have considered here. Some studies report very good agreement between observations and the predictions of $\Lambda$CDM (e.g. Natarajan et~al. 2016) whereas others report a deficit of subhaloes in clusters of a dark matter only simulation \citep[e.g.][]{Grillo2015}. However, on these mass scales (i.e. below $10^{12} \, \Msun$), baryonic physics might play a role.
The number and distribution of {\it massive} substructures in clusters poses a challenge to the cold dark matter model. \cite{Munari2016} found that baryonic effects were unable to reconcile the predictions of $\Lambda$CDM with the abundance of substructures with circular velocities above $200 \, {\rm km \, s}^{-1}$ in Abell 2142 at low redshift ($z=0.09$). Here we have extended the comparison with theory to a simulation with a much larger volume and have looked at an even more complex object, Abell 2744. Now with several examples of clusters which are hard to reconcile with the current standard model, the substructure in massive clusters represents a new test of hierarchical models that is not weakened by baryonic physics \citep{Munari2016} or by uncertainty over the nature of the dark matter particle. More work is needed to make an incontrovertible connection between the peaks in the lensing mass maps with structures in the simulated haloes and to assess the impact of using different halo and subhalo finders; both of these objectives require the full particle data from the N-body simulation and is left for future work (Schwinn et~al. in prep).  

\section*{Acknowledgements}
We are very grateful to the anonymous referee for helpful comments and suggestions. JS thanks Alex Smith for helpful discussions and acknowledges receipt of an ERASMUS\centplus\ internship funded by the European Union. CMB acknowledges receipt of a Leverhulme Trust Research Fellowship. PN acknowledges support from NASA via the grant STScI-49723. This work was supported by the Science and Technology Facilities Council [ST/L00075X/1]. We used the DiRAC Data Centric system at Durham University, operated by the Institute for Computational Cosmology on behalf of the STFC DiRAC HPC Facility (www.dirac.ac.uk). This equipment was funded by BIS National E-Infrastructure capital grant ST/K00042X/1, STFC capital grants ST/H008519/1 and ST/K00087X/1, STFC DiRAC Operations grant ST/K003267/1 and Durham University. DiRAC is part of the National E-Infrastructure.
\bibliographystyle{mnras}%
\bibliography{A2744_MXXL}

\bsp	
\label{lastpage}
\end{document}